\documentclass[twocolumn,showpacs,preprintnumbers,amsmath,amssymb,superscriptaddress]{revtex4-2}

\usepackage{epsf}
\usepackage{graphicx}
\usepackage{sidecap}
 \usepackage{soul}
 \usepackage{array}
 \usepackage{amsmath}
 \usepackage{amssymb}
 \usepackage{dcolumn}
 \usepackage{epstopdf}
 \usepackage{bm}
 \usepackage{amsmath}
\usepackage{amssymb}
\usepackage{amsmath}
\usepackage{mathptmx}
 
\usepackage{gensymb}
\usepackage{color}
\usepackage{hyperref}
\usepackage{soul}
\sethlcolor{green}
\usepackage{bbding}
\hypersetup{
    colorlinks=true,
    citecolor=red,
    linkcolor=red,
    filecolor=blue,   
    urlcolor=blue,
}
 
\def \beq {\begin{equation}}
\def \eeq {\end{equation}}
\renewcommand{\figurename}{{Fig.}}
\renewcommand{\thefigure}{{{\arabic{figure}}}}

\def\bibsection{\refname}
\renewcommand{\refname}{\noindent\textbf{References}}

\begin{document}

\title{{Complex electronic topography and magnetotransport in an in-plane ferromagnetic kagome metal}}

\author{Anup Pradhan Sakhya} \affiliation{Department of Physics, University of Central Florida, Orlando, Florida 32816, USA} 
\affiliation{Research Institute for Synchrotron Radiation Science (HiSOR), Hiroshima University, Higashi-Hiroshima 739-0046, Japan}
\author{Richa Pokharel Madhogaria} \affiliation{Department of Materials Science and Engineering, The University of Tennessee, Knoxville, Tennessee 37996, USA}
\author{Barun Ghosh} \affiliation{Department of Physics, Northeastern University, Boston, Massachusetts 02115, USA}
\affiliation{Quantum Materials and Sensing Institute, Northeastern University, Burlington, Massachusetts 01803, USA}
\affiliation{Department of Condensed Matter and Materials Physics, S. N. Bose National Centre for Basic Sciences, Kolkata-700106, India}
\author{Nabil Atlam} \affiliation{Department of Physics, Northeastern University, Boston, Massachusetts 02115, USA}
\affiliation{Quantum Materials and Sensing Institute, Northeastern University, Burlington, Massachusetts 01803, USA}
\author{Milo Sprague} \affiliation{Department of Physics, University of Central Florida, Orlando, Florida 32816, USA} 
\author{Mazharul Islam Mondal} \affiliation{Department of Physics, University of Central Florida, Orlando, Florida 32816, USA} 
\author{Himanshu Sheokand} \affiliation{Department of Physics, University of Central Florida, Orlando, Florida 32816, USA} 
\author{Arun K. Kumay} \affiliation{Department of Physics, University of Central Florida, Orlando, Florida 32816, USA} 
\author{Shirin Mozaffari} \affiliation{Department of Materials Science and Engineering, The University of Tennessee, Knoxville, Tennessee 37996, USA}
\author{Rui Xue} \affiliation{Department of Materials Science and Engineering, The University of Tennessee, Knoxville, Tennessee 37996, USA}
\author{Yong P. Chen} \affiliation{Department of Physics and Astronomy and Purdue Quantum Science and Engineering Institute,
Purdue University, West Lafayette, Indiana 47907, USA}
\author{David G. Mandrus} \affiliation{Department of Materials Science and Engineering, The University of Tennessee, Knoxville, Tennessee 37996, USA}
\affiliation {Department of Physics and Astronomy, University of Tennessee Knoxville, Knoxville, Tennessee 37996, USA}
\affiliation{Materials Science and Technology Division, Oak Ridge National Laboratory, Oak Ridge, Tennessee 37831, USA}
\author{Arun Bansil} \affiliation{Department of Physics, Northeastern University, Boston, Massachusetts 02115, USA}
\affiliation{Quantum Materials and Sensing Institute, Northeastern University, Burlington, Massachusetts 01803, USA}
\author{Madhab Neupane} \thanks{Corresponding author:\href{mailto:madhab.neupane@ucf.edu}{madhab.neupane@ucf.edu}}\affiliation{Department of Physics, University of Central Florida, Orlando, Florida 32816, USA}

%\author{Anup Pradhan Sakhya et al.}
%\affiliation {Department of Physics, University of Central Florida, Orlando, Florida 32816, USA}
%\author{Madhab~Neupane} \thanks{corresponding author: Madhab.Neupane@ucf.edu} \affiliation {Department of Physics, University of Central Florida, Orlando, Florida 32816, USA}
\date{\today}

\begin{abstract}

\indent The intricate interplay between flat bands, Dirac cones, and magnetism in kagome materials has recently attracted significant attention from materials scientists, particularly in compounds belonging to the \textit{R}Mn$_6$Sn$_6$ family (R = Sc, Y, rare-earths), due to their inherent magnetic frustration. Here, we present a detailed investigation of the ferromagnetic (FM) kagome magnet ScMn$_6$(Sn$_{0.78}$Ga$_{0.22}$)$_{6}$ using angle-resolved photoemission spectroscopy (ARPES), magnetotransport measurements, and density functional theory (DFT) calculations. Our findings reveal a paramagnetic-to-FM transition at 375 K, with the in-plane direction serving as the easy magnetization axis. Notably, ARPES measurements reveal a Dirac cone near the Fermi energy, while the Hall resistivity exhibits a substantial contribution from the anomalous Hall effect. Additionally, we observe a flat band spanning a substantial portion of the Brillouin zone, arising from the destructive interference of wave functions in the Mn kagome lattice. Theoretical calculations reveal that the gap in the Dirac cone can be modulated by altering the orientation of the magnetic moment. An out-of-plane orientation produces a gap of approximately $\sim$ 15 meV, while an in-plane alignment leads to a gapless state, as corroborated by ARPES measurements. This comprehensive analysis provides valuable insights into the electronic structure of magnetic kagome materials and paves the way for exploring novel topological phases in this material class.
%Polarization-dependent ARPES measurements  further elucidate that the Dirac cone originate from Mn \textcolor{blue} {\textit{$d_{x^{2}-y^{2}}$}, \textit{$d_{z^{2}}$}, and \textit{d$_{yz}$}} orbitals, respectively.
\end{abstract}

\maketitle
%\noindent \textbf{Introduction}
\indent Kagome materials have generated considerable excitement in recent years due to their manifestation of various novel topological and correlated electronic phenomena \cite{Han, Guo, Lin, Hasan, Ye, Yin, ghimire, Kang, KangCoSn, LiYMn6Sn6, Sabin, Sabin1, Cheng, Anup, AnupLa, AnupCe, AnupNd}. The kagome lattice consists of a two-dimensional honeycomb network characterized by alternating, corner-sharing triangles that form a hexagonal structure. Analysis using a simple tight-binding model has revealed a band structure featuring Dirac cones, Van Hove singularities, and flat bands, all emerging from this unique lattice geometry \cite{KangCoSn}. These flat bands offer a versatile platform for exploring exotic correlated electron phenomena, where interaction energy dominates over kinetic energy \cite{KangCoSn}. Incorporating spin-orbit coupling (SOC) and magnetism can open a gap in the Dirac band at the corners of the Brillouin zone (BZ), leading to the emergence of intrinsic Chern quantum phases. This, in turn, results in a significant intrinsic anomalous Hall effect (AHE) arising from the Berry curvature of Dirac cones \cite{Yin, Ye, Tang, Lian, Xiao}. Recent discoveries have uncovered several kagome magnets exhibiting exotic topological states and other intriguing phenomena. Notable examples include massive Dirac fermions in Fe$_3$Sn$_2$ \cite{Ye}, a giant intrinsic anomalous Hall effect in the antiferromagnets Mn$_3$Sn and Mn$_3$Ge, and the identification of the magnetic Weyl semimetal Co$_3$Sn$_2$S$_2$ \cite{Nakatsuji, Nayak, Liu}.

The \textit{R}Mn$_6$Sn$_6$ family (where \textit{R} represents rare-earth elements) has recently become a focal point of research \cite{Yin, Ma, Wang, Gu, LiYMn6Sn6, Asaba, Zeng, Y16, Dhakal, Kabir, Lv, LuMn6Sn6}. In LiMn$_6$Sn$_6$, significant contributions to the anomalous Hall effect have been attributed to band crossings near the Fermi energy \cite{chenLi}. Additionally, Chern-gapped massive Dirac fermions have been identified within the \textit{R}Mn$_6$Sn$_6$ family, which features a FM Mn kagome lattice \cite{Ma}. Observations of Landau quantization and the appearance of a Landau fan diagram in TbMn$_6$Sn$_6$ further indicate the existence of massive Dirac fermions at the K point \cite{Yin}. Several members of the \textit{R}Mn$_6$Sn$_6$ family (\textit{R} =Gd-Dy, Y) have been reported to host Dirac fermions and flat bands arising from the Mn magnetic kagome lattice, leading to a significant Berry curvature in momentum space \cite{Wang, Gu, LiYMn6Sn6}. 

Additionally, intrinsic anomalous Hall conductivity has been observed in ferrimagnetic \textit{R}Mn$_6$Sn$_6$ (\textit{R} =Gd-Ho) and helical antiferromagnetic (AFM) YMn$_6$Sn$_6$ materials \cite{Ma, Asaba, Zeng, LGao,WLMa}. Furthermore, a topological Hall effect has been observed in YMn$_6$Sn$_6$, ErMn$_6$Sn$_6$, and HoMn$_6$Sn$_6$, attributed to the complex magnetic structures in these materials \cite{QWang, Y16, Dhakal, Kabir}. As a result, significant attention is being directed towards experimental investigations of various members of the \textit{R}Mn$_6$Sn$_6$ family. The magnetic phases in these materials play a central role in their intriguing properties, necessitating a comprehensive approach to understanding the intricate magnetic interactions and their manifestation in the topological surface states.

\begin{figure}
	\includegraphics[width=8.5cm]{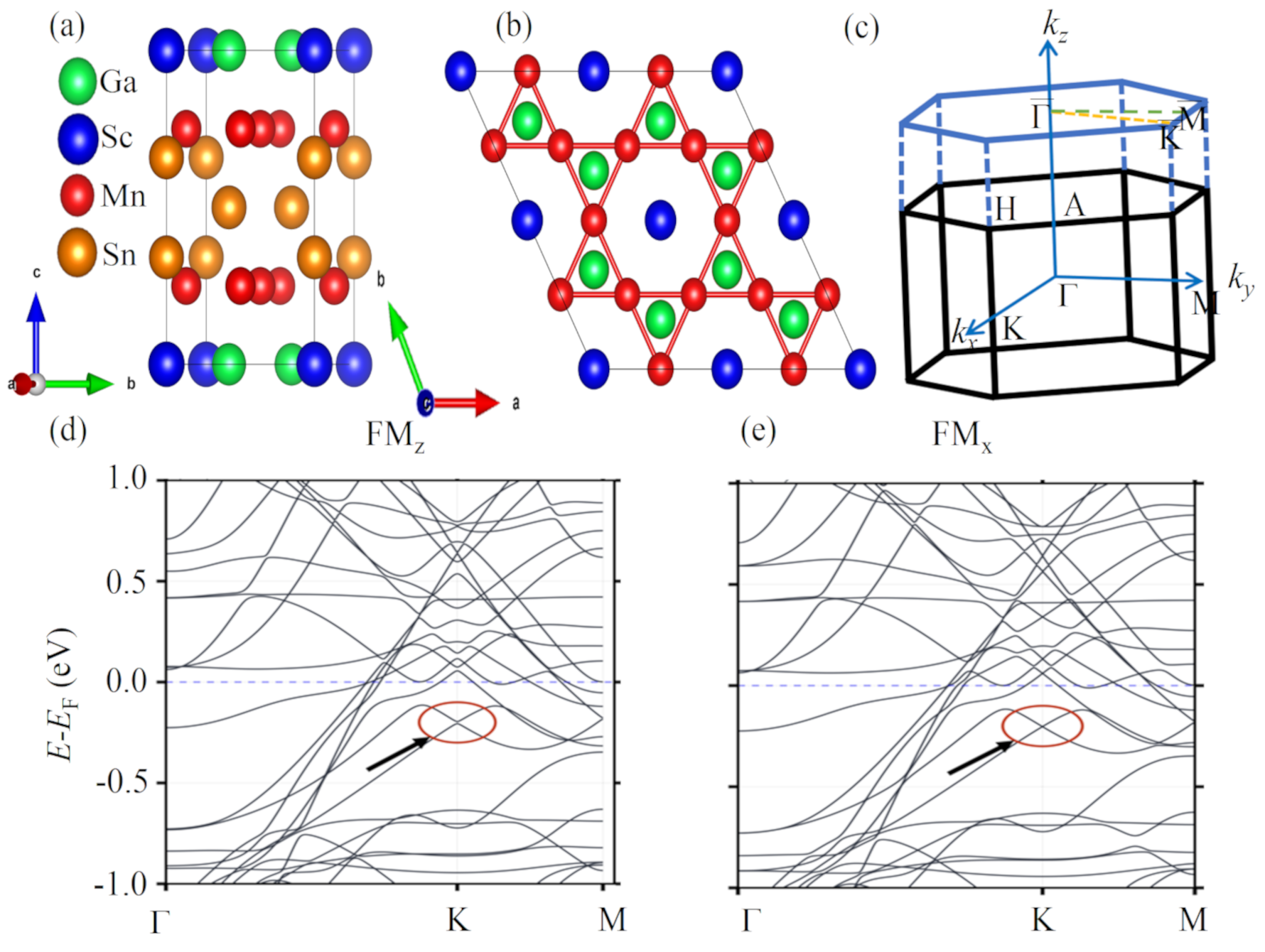}
\centering
\caption{Theoretical modulation of Dirac gap through magnetization direction tuning. (a) Side view and (b) Top view of the crystal structure of ScMn$_6$(Sn$_{0.78}$Ga$_{0.22}$)$_{6}$. Green, Blue, Red, and Orange colored balls represent Ga, Sc, Mn, and Sn atoms, respectively. (c) Bulk BZ and the projection on the (001) surface BZ with marked high-symmetry points. (d) Bulk band structure along the ${\Gamma}$-K-M high symmetry directions calculated with the spin-quantization direction along the [001] and (e) [100] directions, respectively. Spin-orbit coupling is considered in the calculations.} %\textbf{g} Left, middle and right panel presents a schematic showing a nonmagnetic kagome lattice hosting a gapless Dirac cone, a FM kagome lattice with magnetization along the \textit{c}-axis and along the \textit{b}-axis hosting a gapped Dirac cone at the K point of the BZ, respectively. The value of the gap is comparatively much smaller when the magnetization direction is in plane as can be seen from the values shown in the schematic.}
\end{figure}

This study focuses on investigating the influence of spin-orbit interaction and FM ordering on Ga-doped ScMn$_6$Sn$_6$. The undoped ScMn$_6$Sn$_6$ compound exhibits semimetallic behavior and has a hexagonal crystal structure, isostructural to HfGe$_6$Sn$_6$, within the symmetry group \textit{P6/mmm} (No. 191). Its lattice parameters are characterized by a = b= 5.465 \AA\ and c = 8.96 \AA. This layered material consists of alternating Mn and Sn layers \cite{Malaman, Venturini}. The Mn layers form an electronically isolated two-dimensional kagome lattice, while one of the Sn layer adopts a hexagonal crystal structure. Pristine ScMn$_6$Sn$_6$ is composed of double-layered manganese kagome sheets, separated by alternating Sn and Sc-Sn sublayers. In these kagome sheets, the magnetic spins align ferromagnetically within the plane and are connected helically along the \textit{c}-axis \cite{HZhang}. Below the N\'eel temperature (T$_N$) of 390 K, ScMn$_6$Sn$_6$ undergoes AFM ordering, exhibiting multiple magnetic transitions with increasing external magnetic field \cite{Richa}. This family of compounds displays various magnetic phases, including ferromagnetism, ferrimagnetism, antiferromagnetism, and helical ordering, reflecting the complex competition between magnetic phases. Consequently, the delicate magnetic phases in ScMn$_6$Sn$_6$ can be modulated by doping at the Sn sites. Gallium doping stabilizes the FM phase. In this study, ScMn$_6$Sn$_6$ crystals were doped with Ga at the Sn sites, contributing to this stabilization. At lower doping levels (5\%), the helical phase extends slightly to higher temperatures, increasing from 370 K in the parent compound to 417 K. However, as the doping concentration reaches 22\%, intraplanar AFM correlations are suppressed, resulting in a transition from paramagnetic (PM) to FM behavior below 370 K. Our research focuses on this FM phase. 

%By utilizing angle-resolved photoemission spectroscopy (ARPES), we have measured the electronic structure of this compound, which has been supported by first-principles calculations. We have performed systematic studies of transport and magnetic behaviors of the ScMn$_6$(Sn$_{0.78}$Ga$_{0.22}$)$_{6}$  kagome magnet.

\begin{figure*}
	\centering
	\includegraphics[width=18cm]{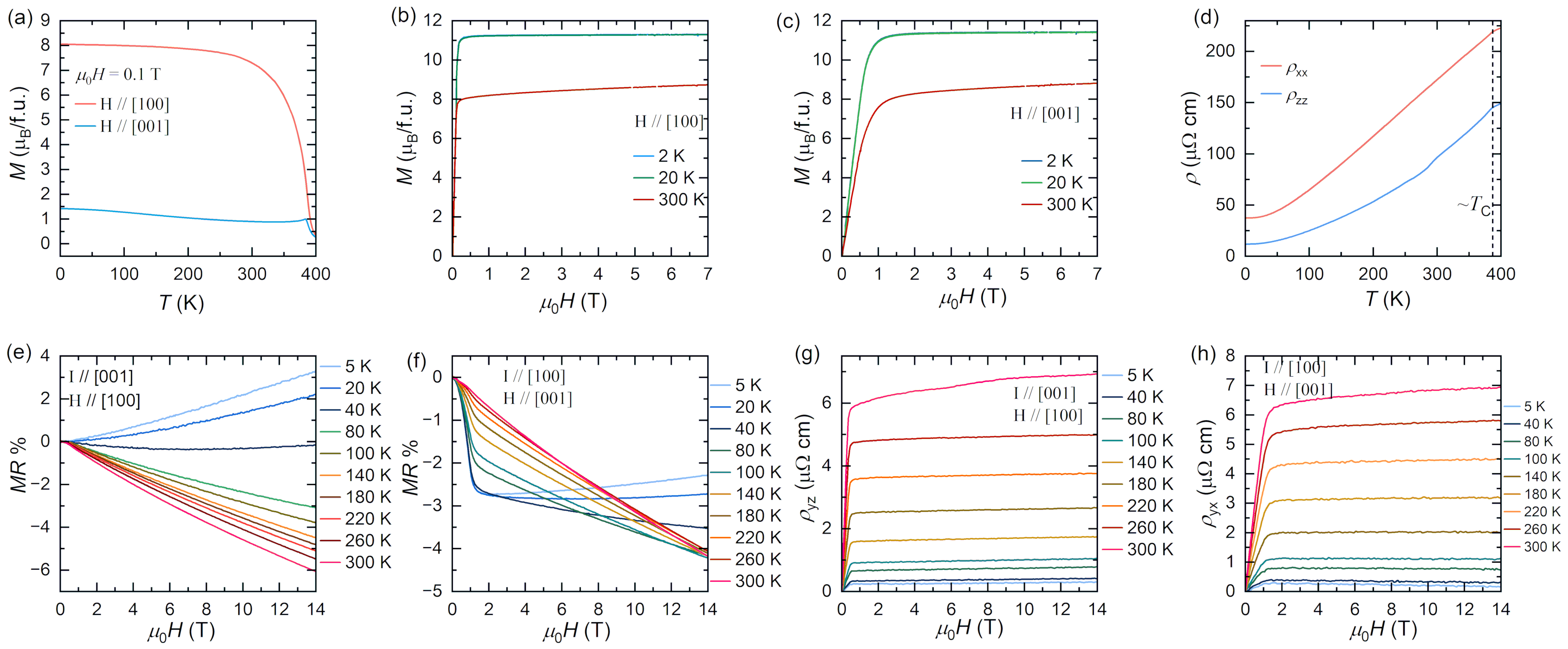}
\vspace{-1ex}
	\caption{Observation of in-plane ferromagnetization and the anomalous Hall effect. (a) Temperature dependence of magnetization with an applied magnetic field along the \textit{ab} plane and the \textit{c}-axis. (b) Isothermal magnetization curves along the \textit{ab} plane and the (c) \textit{c}-axis as a function of temperature. (d) Temperature dependence of the longitudinal resistivity along the \textit{ab} plane and the \textit{c}-axis. (e) Magnetoresistance for current along the \textit{c}-axis, \textit{H} is applied in the \textit{ab} plane. (f) Magnetoresistance for current along the \textit{ab}-plane, \textit{H} is applied in the \textit{c}-axis. (g) Hall resistivity along the \textit{ab}-plane, \textit{I} is applied along the \textit{c}-axis. (h) Hall resistivity along the \textit{c}-axis, \textit{I} is applied along the \textit{ab}-plane.}
\end{figure*}

In this Letter, we present an in-depth investigation into the transport, magneto-transport and the electronic structure of the kagome magnet ScMn$_6$(Sn$_{0.78}$Ga$_{0.22}$)$_{6}$ in its FM state. Our methodology combines angle-resolved photoemission spectroscopy (ARPES) with density functional theory (DFT) calculations. A clear transition from paramagnetic to FM behavior is observed around \textit{T$_c$} =375 K in ScMn$_6$(Sn$_{0.78}$Ga$_{0.22}$)$_{6}$. This system exhibits strong magnetic anisotropy, favoring an in-plane easy axis. The Hall resistivity indicates substantial contribution from the anomalous Hall effect. 
Additionally, we identified a Dirac cone at the $\overline{{\text{K}}}$ point with a binding energy of approximately 0.1 eV below the Fermi level (E$_F$). We also observed a flat band extending across a significant portion of the BZ, originating from the destructive interference of the wave function within the Mn kagome lattice. Theoretical calculations reveal the Dirac cone gap can be tuned by changing the magnetic moment orientation where an out-of-plane alignment creates a $\sim$ 15 meV gap, while an in-plane orientation results in a gapless state, as confirmed by ARPES. Although strong correlations, non-trivial band topology, magnetic ordering, and geometric frustration have each been extensively studied individually, it is rare for a material family to exhibit multiple, if not all, of these characteristics simultaneously. Our study sheds light on the multifaceted nature of this material, opening new avenues for exploring its novel properties.

\indent High-quality single crystals of ScMn$_6$(Sn$_{0.78}$Ga$_{0.22}$)$_{6}$ were synthesized by the self-flux method. Magnetization measurements were carried out using the vibrating sample magnetometer (VSM). Resistivity and Hall measurements were performed using the conventional four-probe method. Synchrotron based ARPES measurements were performed at the Advanced Light Source (ALS) beamline 4.0.3 equipped with R8000 hemispherical analyzer. We performed density functional theory (DFT) based ab initio calculations using the Vienna Ab Initio Simulation Package (VASP) with Projector Augmented Wave (PAW) pseudopotentials and a plane-wave basis set \cite{Kresse1, Kresse2}. Details on the experimental and computational methods have been provided in the Supplemental Material (SM) section 1 \cite{SI}.   

\indent ScMn$_6$(Sn$_{0.78}$Ga$_{0.22}$)$_{6}$ crystallizes in the hexagonal MgFe$_6$Ge$_6$-type structure with space-group \textit{P6/mmm} (No. 191), as shown in Fig. 1(a-b). Within a unit cell, two pristine Mn kagome layers exhibit a simple A-A stacking pattern, while the hexagonal structure is formed by Sn and Ga atoms, with Sc atoms positioned at the center of the hexagon. The magnetic configuration features FM Mn planes, and the kagome lattice formed by Mn atoms is illustrated in Fig. 1(b). Rietveld refinement of room temperature powder X-ray diffraction (PXRD) data using Fullprof software yielded lattice parameters of a = b = 5.409 \AA\ and c= 8.846 \AA\ (see SM section 2 and Fig. S1) \cite{SI}. Figure 1(c) shows the bulk 3D BZ with high-symmetry points and the projected 2D BZ along the \textit{c}-axis. Experimental lattice constants were used for the present DFT calculations, with the SOC calculated FM bulk band structure along the $\Gamma$--$\text{K}$--$\text{M}$ direction shown in Fig. 1(d) and Fig. 1(e), with spin-quantization directions along the out-of-plane and in-plane orientations, respectively. The Dirac cone at the $\text{K}$-point is evident in both figures. Notably, when the spin-quantization direction is oriented out-of-plane, a small but finite gap opens in the Dirac cone, which becomes gapless when the spin-quantization direction lies in-plane. 

%Figure 1(g) illustrates a schematic depicting the variation in the size of the SOC-induced gap for both non-magnetic and magnetic systems (present case), with magnetic moments aligned along different directions. In a non-magnetic system, there is no SOC-induced gap at the $\text{K}$-point. However, for a magnetic system, the gap size is greater when the magnetic moment aligns along the (001) direction compared to when it aligns along the in-plane direction (100). This indicates the potential to manipulate the SOC gap size by altering the spin-quantization direction. 
To gain a deeper understanding of how the Dirac cone gap varies with the orientation of the magnetization axis, we employ a tight-binding model for the kagome lattice that incorporates a Kane-Mele SOC term. We observe that the Dirac gap size decreases as the magnetic moment rotates away from the $z-$ axis (see Fig. S2 in the SM \cite{SI}). This gap reduction is attributed to the weakened influence of the Kane-Mele SOC as the magnetic moment aligns in the in-plane direction. To illustrate this, consider the following model for spin-polarized electrons in real space \cite{Tang}: 
\begin{equation}
    H(\vec{k}) = t \sum_{\langle i, j\rangle} c^{\dagger}_{i}c_{j} + i \lambda \sum_{\langle i, j\rangle} \vec{S}\cdot (\vec{E}_{ij} \times \vec{\bold{R}}_{ij}) \ c^{\dagger}_{i}  \ c_{j} + \text{h.c}
\end{equation}

The second term represents the Kane-Mele SOC. Here, $\vec{\bold{R}_{ij}}$ is the distance vector between sites $i$ and $j$, and $\vec{E}_{ij}$ denotes the electric field experienced by electrons along $\vec{R}_{ij}$. Assuming the magnetization axis forms an angle $\theta$ with respect to the $\Hat{z}$ axis, the last term can be expressed as: 

\begin{equation}
    H_{\text{KM}} = i \lambda  S \cos{\theta} \sum_{\langle i, j\rangle} \Hat{z}\cdot (\vec{E}_{ij} \times \vec{\bold{R}}_{ij}) \ c^{\dagger}_{i} \ c_{j} + \text{h.c}
\end{equation}

According to this expression, the strength of the Kane-Mele term reaches its maximum when the magnetization axis aligns along the $z$-direction, and approaches zero as the magnetization axis rotates toward the $xy$ plane. This results in a reduced band gap in the FM$-x$ phase compared to the FM$-z$ phase. We note that, in real materials, the SOC may be more complex than the simplified Kane-Mele model used here. Nonetheless, this model provides an intuitive understanding of the band gap dependence at the Dirac point with respect to magnetic orientations. 

\begin{figure*}
\includegraphics[width=16cm]{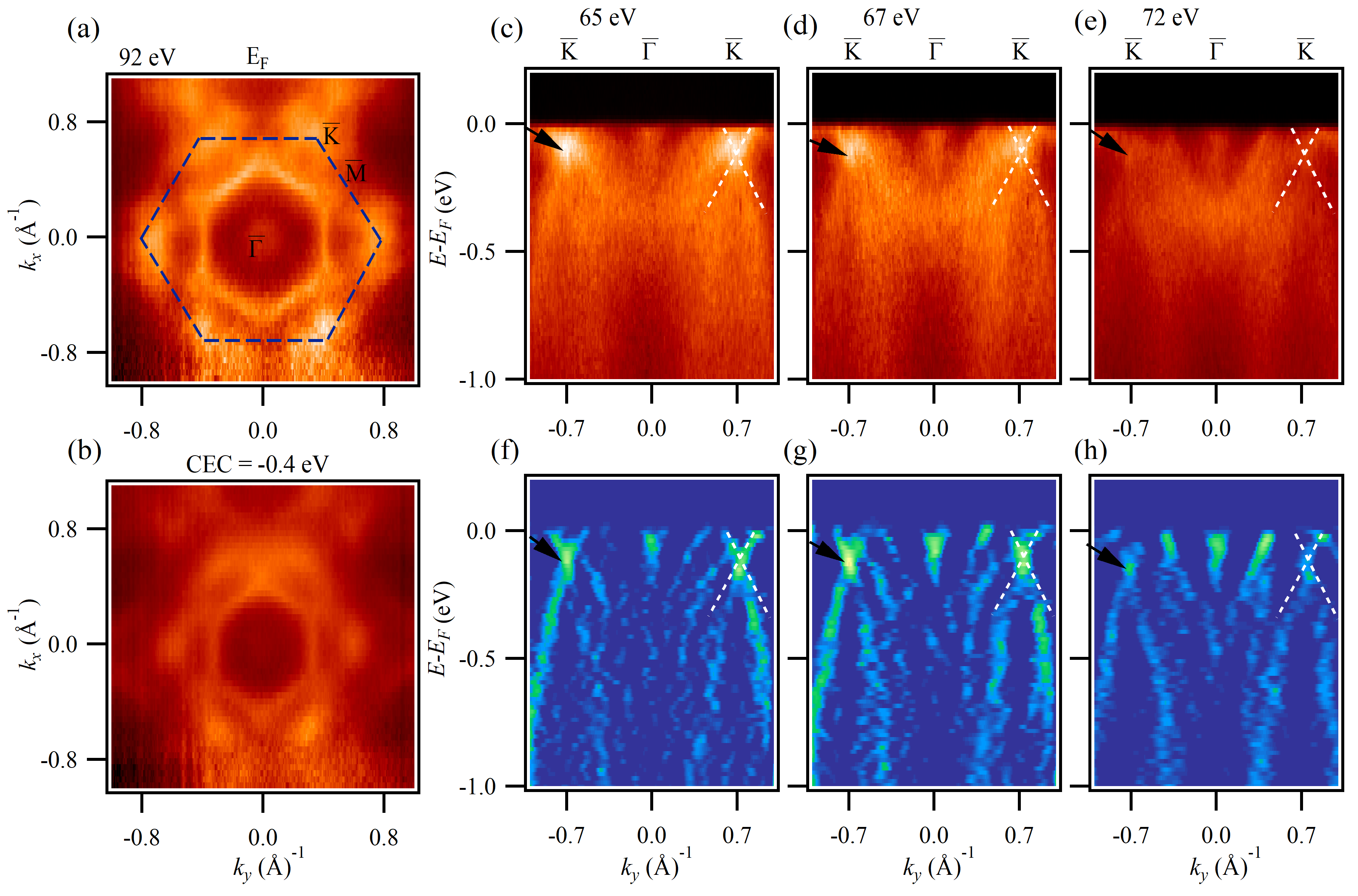}
\caption{Observation of Dirac cone at the $\overline{\text{K}}$ point of the Brillouin zone. (a) ARPES measured Fermi surface (FS) measured along the (001) direction using a photon energy of 92 eV and linear horizontal (LH) polarization. High-symmetry points are labeled in black color. (b) Constant energy contour at a binding energy of - 400 meV. ARPES measured band dispersion along the $\overline{\text{K}}$--$\overline{\Gamma}$--$\overline{\text{K}}$ high symmetry line at a photon energy of (c) 65 eV, (d) 67 eV, and (e) 72 eV, respectively. (f-h) Second derivative plot of the ARPES-measured band dispersion corresponding to panels (c-e), respectively. ARPES measurements were performed at the ALS beamline 4.0.3 at a temperature of 11 K i.e, in the FM phase using LH polarization.}
\end{figure*}

Figure 2(a) shows the temperature-dependent magnetization curves along the \textit{ab} plane and the \textit{c}-axis at $\mu_0H$ = 0.1 T, indicating a transition from PM to FM behavior around T$_c$ = 375 K in ScMn$_6$(Sn$_{0.78}$Ga$_{0.22}$)$_{6}$. This system exhibits strong magnetic anisotropy, favoring an in-plane easy axis. Notably, the in-plane magnetization (red curve) surpasses the out-of-plane susceptibility (blue curve) by approximately fivefold, as shown in Fig. 2(a). Figures 2(b) and 2(c) present isothermal magnetization curves for magnetic fields parallel to the \textit{ab} plane and the \textit{c}-axis at various temperatures. At 2 K, the magnetizations along the \textit{ab} plane and the \textit{c}-axis reach saturation around 0.05 T and 1 T, respectively. Although both directions achieve a similar saturated magnetization ($\approx 11 \mu_B$/f.u.), saturation occurs more quickly along the \textit{ab} plane, confirming it as the preferred direction for magnetization. 

The temperature dependence of the longitudinal resistivity, with current flowing along the \textit{ab} plane ($\rho_{xx}$) and the \textit{c}-axis ($\rho_{zz}$), exhibits typical metallic behavior as temperature decreases from 400 K to 1.8 K. Both resistivity curves show a slight kink at 385 K, suggesting reduced charge-carrier scattering near the PM--FM ordering temperature. The magnetoresistance (MR) for $\mu_0H$ = 0 -- 14 T with $I \parallel$ [001], $H \parallel$ [100] and $I \parallel$ [100], $H \parallel$ [001], is shown in Fig. 2(e) and Fig. 2(f) at selected temperatures. MR was calculated from resistivity data using the formula MR =($\rho_{xx}(\mu_0H)-\rho_{xx})/\rho_{xx}\times100\%$, where $\rho_{xx}$ and $\rho_{xx}(\mu_0H)$ represent zero-field and in-field resistivities, respectively. When $I$ is along the $c$-axis and $H$ is along the $a$-axis [Fig. 2(e)], the MR exhibits nonlinearity across all measured temperatures. For $T \geq 40 K$, MR is negative, indicating reduced scattering due to spin disorder in the FM state at higher temperatures. However, for $T \leq 20 K$, MR becomes positive, suggesting the presence of a different mechanism at low temperatures. Conversely, MR for $I \parallel$ [100], $H \parallel$ [001] remains negative across all temperatures [Fig. 2(f)]. Notably, the slope of the MR curves steepens at lower fields as the temperature decreases. Fig. S3 in the SM \cite{SI} provides a schematic of the crystallographic axes along which the current, external magnetic field, and voltage are applied. While crystal misalignment can cause negative MR, the crystal in this study was carefully aligned, as indicated in the individual panels of Fig. 2. Therefore, the observed negative MR arises solely from the ferromagnetic nature of the system. To investigate the anomalous Hall effect, the Hall resistivity was measured from 5-300 K. Figures 2(g) and 2(h) show the Hall resistivities measured in two different crystallographic directions. In Fig. 2(g), $\rho_{yz}$ represents $I \parallel$ [001], $H \parallel$ [100], while in Fig. 2(h), $\rho_{yx}$ is for $I \parallel$ [100], $H \parallel$ [001]. The Hall resistivity behavior closely resembles the magnetization data M(H) shown in Fig. 2(b) and Fig. 2(c), indicating a substantial contribution from the anomalous Hall effect (AHE). Furthermore, at lower temperatures the observation of positive and negative slope in $\rho_{yz}$ ($H \parallel$ [100]) and  $\rho_{yx}$ ($H \parallel$ [001]), respectively, represents the dominance of holes (electrons) when $H \parallel$ [100], $H \parallel$ [001]. This suggests simultaneous presence of both charge carriers in our compound which may account for the unsaturated positive MR at $T \le 20$ K [Fig. 2(e)] \cite{chenLi}.

Next, to investigate the electronic structure of ScMn$_6$(Sn$_{0.78}$Ga$_{0.22}$)$_{6}$, we performed high-resolution ARPES measurements on the cleaved (001) plane. The results of these measurements are shown in Figures 3-4. Figures 3(a-b) present the Fermi surface (FS) and the constant energy contour (CECs) measured with a photon energy of 92 eV and LH polarized light at a temperature of 11 K, within the FM phase; see SM section 3 and Fig. S4 for additional FS data \cite{SI}. The FS maps reveal multiple pockets at the Fermi level (E$_F$), consistent with the metallic character of the material. In line with the crystal symmetry, the FS displays hexagonal symmetry. Notably, the FS includes a small circular pocket at the $\overline{\Gamma}$ high-symmetry point, while along the $\overline{\Gamma}$--$\overline{{\text{M}}}$/$\overline{{\text{K}}}$ high-symmetry line, a hexagonal pocket emerges. These pockets gradually diminish with increasing binding energy, indicating their electron-like nature. Additionally, intense circular features appear at the BZ corners, corresponding to the $\overline{{\text{K}}}$ points. The DFT-calculated bulk FS and CEC are shown in Fig. S5; see SM Section 3 for details \cite{SI}, demonstrating a reasonable match to the observed ARPES spectra.

Subsequently, we analyzed the band dispersion along various high-symmetry directions as shown in Figs. 3(c-h) and 4. Figure 3(c) illustrates the ARPES-measured band-dispersion along the $\overline{{\text{K}}}$--$\overline{\Gamma}$--$\overline{{\text{K}}}$ high-symmetry line, captured using a photon energy of 65 eV. A small electron-like pocket is visible at the $\overline{\Gamma}$ point, while the spectra reveal a Dirac cone at the $\overline{{\text{K}}}$ point, with the Dirac point located at approximately \textit{E-E$_F$} = $\sim$ -0.1 eV, as indicated by the white dashed lines for clarity. Measurements conducted with photon energies of 67 eV and 72 eV, shown in Figs. 3(d-e), demonstrate that while the intensity of the band forming the Dirac cone vary, the band remain distinctly visible. The Dirac cone at the $\overline{{\text{K}}}$ point is more clearly visualized in the second-derivative plots in Figs. 3(f-h), where it is highlighted by white-dashed lines and black arrows for clarity. The DFT-calculated band dispersion along the $\overline{{\text{K}}}$--$\overline{\Gamma}$--$\overline{{\text{K}}}$ line, presented in Fig. S6, clearly shows the Dirac cone (see SM section 4 for further details \cite{SI}). Its appearance in bulk DFT calculations indicates a bulk origin for the Dirac cone.

%The orbital character DFT calculated plots along the $\overline{{K}}$--$\overline{\Gamma}$--$\overline{{K}}$ high-symmetry line, presented in Fig. 4(f) suggests this Dirac cone originates from the Mn kagome lattice with significant contribution from Mn 3\textit{d} electrons.  %The band-dispersion along the $\overline{\Gamma}$--$\overline{\textit{K}}$--$\overline{\textit{M}}$--$\overline{\textit{K}}$--$\overline{\Gamma}$ high-symmetry line is presented for a photon energy of 114 eV and 128 eV which corresponds to \textit{k$_z$} = 0 and \textit{k$_z$} = $\pi$ plane, respectively. Similar electronic features in both Fig. 4(a) and (b) suggests negligible dispersion of the observed bands. The observed bands are found to be weakly dispersing in \textit{k$_z$}, demonstrating the isolated 2D electronic structure of the manganese kagome lattice. 

%\begin{figure*}
%\centering
%\includegraphics[width=16cm]{Figure5.png}
%\caption{Photon-energy-dependent band structure along the $\overline{\Gamma}$--$\overline{\text{K}}$--$\overline{\text{M}}$--$\overline{\text{K}}$--$\overline{\Gamma}$ high-symmetry line. (a-b) ARPES measured band structure along the $\overline{\Gamma}$--$\overline{\text{K}}$--$\overline{\text{M}}$--$\overline{\text{K}}$--$\overline{\Gamma}$ high-symmetry line at various photon energies as indicated on top of the plots. ARPES measurements were performed at the ALS beamline 4.0.3 at a temperature of 11 K i.e, in the ferromagnetic phase using LH polarization.}
%\end{figure*} 

\begin{figure}
\centering
\includegraphics[width=8.5cm]{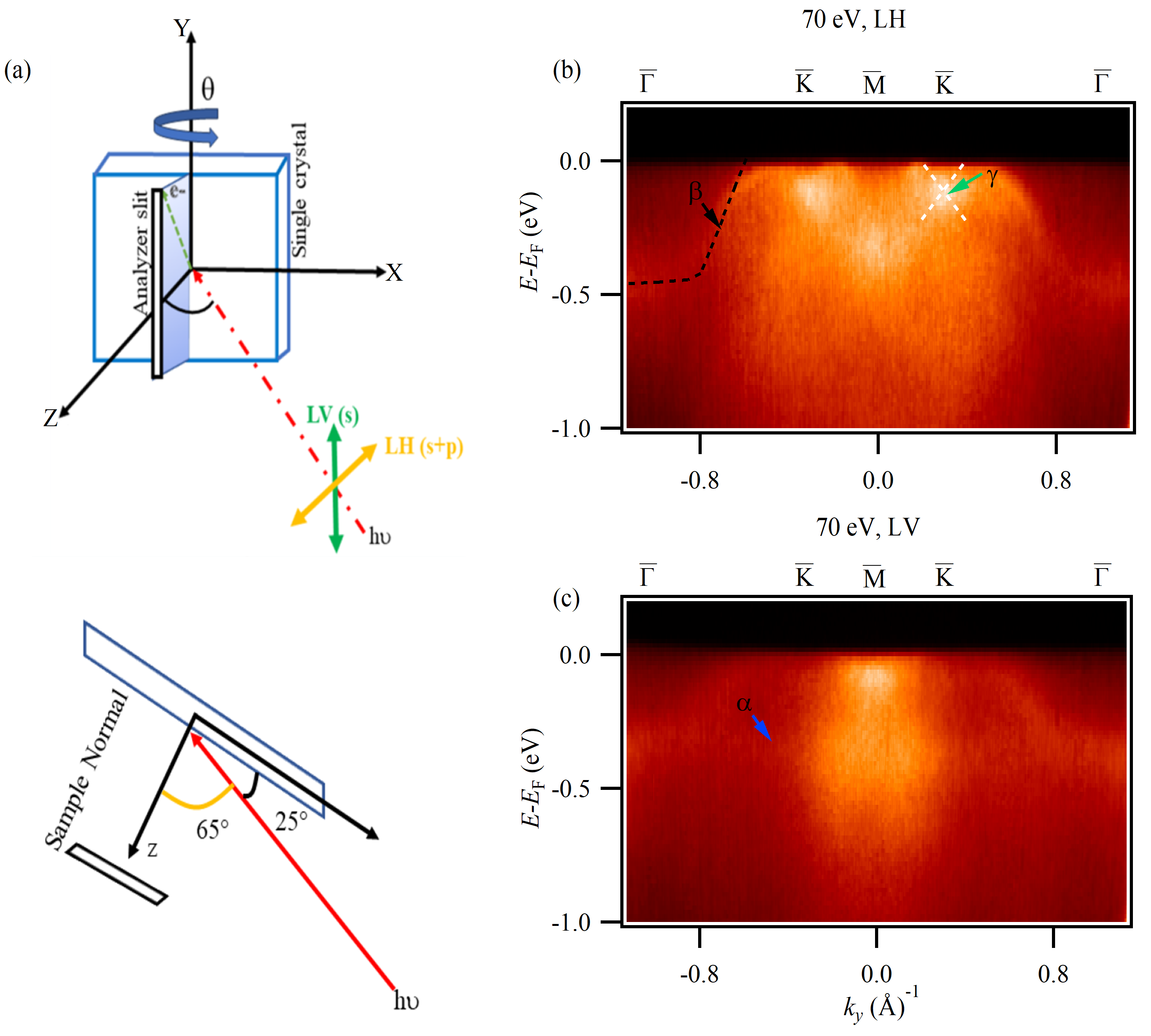}
\caption{Polarization-dependent behavior of the flat band and Dirac cone. (a) Experimental geometry used for polarization-dependent ARPES measurements. Experimental band dispersion measured along the $\overline{\Gamma}$--$\overline{\text{K}}$--$\overline{\text{M}}$--$\overline{\text{K}}$--$\overline{\Gamma}$ high-symmetry line using (b) LH polarization and (c) LV polarization. The symbols $\alpha$, $\beta$, and $\gamma$ represent the flat band, parabolic band, and Dirac cone, respectively. ARPES measurements were performed at the ALS beamline 4.0.3 at a temperature of 11 K i.e, in the FM phase using both LH and LV polarization.}
\end{figure}

To gain further insights into the orbital character of the electronic bands, we conducted polarization-dependent ARPES measurements. Figure 4(a) shows the experimental setup, where measurements were performed along the $\overline{\Gamma}$--$\overline{\text{K}}$--$\overline{\text{M}}$--$\overline{\text{K}}$--$\overline{\Gamma}$ high-symmetry line using linear horizontal (LH) and linear vertical (LV) polarizations, as depicted in Fig. 4(b-c), respectively. As evident from the figure, different sets of bands are selectively enhanced under LH and LV polarized light, respectively, revealing interesting distinctions between the two data sets. A flat band with a bandwidth of $\sim$ 40 meV denoted as $\alpha$ at approximately $\sim$ -0.4 eV binding energy is detected with LV polarization but not with LH, suggesting these features arise from different orbitals. A dispersive band, labeled $\beta$, crosses the E$_F$ and extends from the $\overline{\text{K}}$ to the $\overline{\Gamma}$ high-symmetry line. Although this band is slightly more intense with LH polarization, it remains visible in both LH and LV polarizations. Notably, a Dirac cone appears at the $\overline{\text{K}}$ point under LH polarization but is absent under LV polarization (see SM section 4-5 and Fig. S7-S9 for additional photon energy dependent ARPES measurements along the $\overline{\Gamma}$--$\overline{\text{K}}$--$\overline{\text{M}}$--$\overline{\text{K}}$--$\overline{\Gamma}$ line \cite{SI}). Following the selection rules in photoemission, the ability to selectively excite bands depends on their orbital symmetry relative to specific mirror planes within the setup. Specifically, polarizations with the electric field vector aligned either parallel or perpendicular to the mirror plane distinguish orbitals with even or odd parity concerning the mirror plane. When aligning the $\overline{\Gamma}$--$\overline{\text{K}}$--$\overline{\text{M}}$--$\overline{\text{K}}$--$\overline{\Gamma}$ high-symmetry line along the analyzer slit (i.e., along the \textit{y}-\textit{z} mirror plane), LV polarization suppresses odd orbitals and detects even orbitals. Based on selection rules, the $\alpha$ band is attributed to Mn \textit{$d_{x^{2}-y^{2}}$}, \textit{$d_{z^{2}}$}, and \textit{d$_{yz}$} orbitals. Using similar arguments, the $\gamma$ band originates from Mn \textit{d$_{xy}$} and Mn \textit{d$_{xz}$} orbitals.

%Notably, there is a small gap opening in the Dirac cone from the DFT calculations, and the calculated chirality of the bands suggests it to be topological in nature. Significant anomalous Hall effect in this material is potentially due to the Berry curvature contribution from this gapped Dirac cone. \\ 
\indent While no ARPES studies have been reported on the parent material ScMn$_6$Sn$_{6}$ or ScMn$_6$(Sn$_{0.78}$Ga$_{0.22}$)$_{6}$, some Mn-based kagome materials have been investigated using ARPES. Magnetic measurements on YMn$_6$Sn$_{6}$ reveal two transition temperatures: T$_N$ = 359 K, corresponding to a paramagnetic-to-AFM transition, and T$_{hel}$ = 326 K, where a \textit{c}-axis helical order with in-plane ferromagnetism emerges \cite{Y16}. ARPES studies on YMn$_6$Sn$_{6}$ have shown that it hosts a Dirac cone, with the Dirac point located 45 meV below the E$_F$ in the AFM phase \cite{Y16}. In contrast, TbMn$_6$Sn$_{6}$ exhibits a ferrimagnetic ground state with a Curie temperature of 423 K \cite{Yin}. Magnetization measurements on TbMn$_6$Sn$_{6}$ show a well-defined hysteresis loop when the field is applied along the \textit{c}-axis but no loop when applied in the \textit{a-b} plane, confirming strong out-of-plane magnetization. Notably, scanning tunnelling microscopy and ARPES measurements on TbMn$_6$Sn$_{6}$ reveal a large Chern gap at 130 meV above E$_F$ in the ferrimagnetic ground state \cite{Yin}. Another material, ErMn$_6$Sn$_{6}$ undergoes a paramagnetic-to-AFM transition at T$_N$ = 345 K, followed by a ferrimagnetic transition below 68 K \cite{Dhakal}. However, ARPES measurements did not reveal clear Dirac cones or flat bands in its ferrimagnetic ground state \cite{Dhakal}. Similarly, HoMn$_6$Sn$_{6}$ transitions from a paramagnetic to a ferrimagnetic state below 376 K and undergoes a spin-reorientation transition below 200 K \cite{Kabir}. ARPES studies on HoMn$_6$Sn$_{6}$ suggest that the Dirac cone lie above E$_F$, while flat bands remain undetected \cite{Kabir}. In comparison, ScMn$_6$(Sn$_{0.78}$Ga$_{0.22}$)$_{6}$ adopts a FM ground state at low temperatures without undergoing additional magnetic transitions, making it an excellent system for studying metallic behavior without the complexities of multiple magnetic phases. The Dirac point in ScMn$_6$(Sn$_{0.78}$Ga$_{0.22}$)$_{6}$ is located approximately 100 meV below E$_F$, slightly lower than in YMn$_6$Sn$_{6}$ (45 meV below E$_F$) \cite{Y16}, while in TbMn$_6$Sn$_{6}$, the Dirac point is positioned 130 meV above E$_F$ \cite{Yin}. The magnetic behavior of ScMn$_6$(Sn$_{0.78}$Ga$_{0.22}$)$_{6}$ contrasts with that of its parent compound, ScMn$_6$Sn$_{6}$, which undergoes AFM ordering below T$_N$ =390 K and transitions through multiple magnetic phases under an external magnetic field \cite{Richa}. A key result of this study is the successful doping of 22 \% Ga at Sn sites in ScMn$_6$Sn$_{6}$, stabilizing ferromagnetism with an in-plane easy axis in ScMn$_6$(Sn$_{0.78}$Ga$_{0.22}$)$_{6}$. The material undergoes a paramagnetic-to-FM transition at T$_c$ = 375 K, distinguishing it from the above discussed Mn-based 166 kagome metals. Notably, the in-plane magnetization of ScMn$_6$(Sn$_{0.78}$Ga$_{0.22}$)$_{6}$ is approximately five times greater than its out-of-plane susceptibility, further highlighting its unique magnetic anisotropy. 

\indent In summary, we have tuned the magnetic properties of ScMn$_6$Sn$_{6}$ by incorporating Ga as dopant, yielding the compound ScMn$_6$(Sn$_{0.78}$Ga$_{0.22}$)$_{6}$, which exhibits FM behavior below 375 K, with an in-plane easy magnetization axis. Hall resistivity measurements reveal a prominent anomalous Hall effect, accompanied by the observation of a Dirac cone near E$_F$ in both ARPES experiments and DFT calculations. Additionally, we identify flat bands extending to a significant section of the BZ, originating from the destructive interference within the Mn kagome lattice. While a flat band is present, its position below the Fermi level suggests that strong electronic correlations do not play a dominant role in this system. In a broader perspective, our research offers new pathways for investigating the diverse and intriguing physics of these Mn-based kagome lattices.

%\noindent \textbf{ACKNOWLEDGMENTS}\\
\indent M.N. acknowledges support from the Air Force Office of Scientific Research MURI (Grant No. FA9550-20-1-0322) and the US Department
of Energy (DOE), Office of Science, Basic Energy Sciences grant number DE-SC0024304. DGM acknowledges the support from AFOSR MURI (Novel Light-Matter Interactions in Topologically Non-Trivial Weyl Semimetal Structures and Systems), grant FA9550-20-1-0322. The work at Northeastern University was supported by the Air Force Office of Scientific Research under Award No. FA9550-20-1-0322, and it benefited from the computational resources of Northeastern University’s Advanced Scientific Computation Center (ASCC) and the Discovery Cluster. This research used resources of the Advanced Light Source, a U.S. Department of Energy Office of Science User Facility, under Contract No. DE-AC02-05CH11231. We thank Jonathan Denlinger for beamline assistance at the Advanced Light Source (ALS), Lawrence Berkeley National Laboratory.\\

\vspace{2ex}

\raggedbottom
\clearpage
\pagebreak{}

%%%%%%%%%% Merge with supplemental materials %%%%%%%%%%
\clearpage
\widetext
\begin{center}
\textbf{\large Supplemental Materials for\\[0.3cm]
Complex electronic topography and magnetotransport in an in-plane ferromagnetic kagome metal}
\end{center}
%%%%%%%%%% Merge with supplemental materials %%%%%%%%%%
%%%%%%%%%% Prefix a "S" to all equations, figures, tables and reset the counter %%%%%%%%%%
\setcounter{equation}{0}
\setcounter{figure}{0}
\setcounter{table}{0}
\setcounter{page}{1}
\makeatletter

% Continue reference numbering from main text
\setcounter{NAT@ctr}{44}

\renewcommand{\theequation}{S\arabic{equation}}
\renewcommand{\figurename}{{Fig.}}
\renewcommand{\thefigure}{{{S\arabic{figure}}}}
\renewcommand{\bibnumfmt}[1]{[#1]}
\renewcommand{\citenumfont}[1]{#1}
\renewcommand{\tablename}{Supplementary Table}
\renewcommand{\thetable}{\arabic{table}}
\def\bibsection{\refname}
\renewcommand{\refname}{\noindent\textbf{Supplementary References}}

\noindent\textbf{1. Methods.}\\
\noindent \textbf{Crystal synthesis.}

\noindent High-quality single crystals of ScMn$_6$(Sn$_{0.78}$Ga$_{0.22}$)$_{6}$ were synthesized by the self-flux method with high-purity Sc(Alfa Aesar 99.9 \%), Ga(Alfa Aesar 99.9 \%), Mn(Alfa Aesar 99.98 \%), Sn shot (Alfa Aesar 99.999 \%). These pieces were taken in the atomic ratio of Sc:Mn:Sn:Ga = 1:6:27:3 in a Canfield crucible set \cite{Canfield} and sealed in an evacuated silica tube under vacuum. The sealed ampoule was heated at 60 \degree C h$^{-1}$ to 973 \degree C, followed by a 48 hour dwell. Crystals were grown during slow cool (1.2 \degree C h$^{-1}$) to 600 \degree C. The ampoule was centrifuged at 600 \degree C to remove the excess Sn flux, hexagonal shaped shiny single crystals were obtained from the synthesis. We have verified the crystal structure using the PXRD method using a PANanalytical Empyrean diffractometer with a Cu-K$\alpha$ source. The chemical composition and homogeneity of the crystals were confirmed through energy dispersive X-ray spectroscopy analysis.\\ 

\noindent \textbf{Magnetization and Transport Measurements.}

\noindent Magnetization measurements were carried out using the vibrating sample magnetometer (VSM) option of a physical property measurement system (PPMS) from Quantum Design. Resistivity and Hall measurements were performed using the conventional four-probe method. Single crystals were shaped into rectangular bar to perform the transport experiments. Magnetoresistance and Hall measurements were carried out such that current (\textit{I}) was along [100]/[001] and magnetic field was applied along [001]/[100] crystallographic direction. Resistivity option of PPMS was used to implement the electrical measurements.\\

\noindent \textbf{Angle-resolved photoemission spectroscopy (ARPES) measurements.}

\noindent Synchrotron based ARPES measurements were performed at the Advanced Light Source (ALS) beamline 4.0.3 equipped with R8000 hemispherical analyzer. In order to obtain high-quality and fresh surfaces required for ARPES experiments, single crystals were cleaved in situ under ultra high vacuum with pressure maintained in the order of 10$^{-11}$ torr. The measurements were carried out at a temperature of 11 K i.e., in the FM phase. LH and LV polarized photon sources in the range of 30 - 124 eV were used during the measurements.\\ 

\noindent \textbf{Computational methods.}\\
\noindent We performed density functional theory (DFT) based ab initio calculations using the Vienna Ab Initio Simulation Package (VASP) with Projector Augmented Wave (PAW) pseudopotentials and a plane-wave basis set \cite{Kresse1, Kresse2}. The kinetic energy cutoff for the plane-wave basis was set to 500 eV, and a 12$\times$12$\times$6 \textit{k}-point mesh was used for Brillouin zone integration. We adopted the generalized gradient approximation (GGA) framework developed by Perdew-Burke-Ernzerhof to treat the exchange-correlation potential \cite{Perdew}. We constructed Wannier function-based \cite{Pizzi} tight-binding models for ScMn$_6$Sn$_6$ and SnMn$_6$Ga$_6$ separately and adopted the virtual crystal approximation (VCA) to model the 22\% Ga-doped structure. The VCA neglects local disorder effects from Ga substitution which may cause band broadening, althought the sharp ARPES features indicate this effect is minimal here. All DFT calculations were performed with the Hubbard \textit{U} parameter set to zero, as introducing a finite \textit{U} leads to theoretical magnetic moments that deviate significantly from the experimentally observed values.\\

\noindent\textbf{2. X-ray diffraction data of ScMn$_6$(Sn$_{0.78}$Ga$_{0.22}$)$_{6}$ and Configuration for Magnetoresistance Measurements.}\\
Figure S1 displays the X-ray diffraction (XRD) data for ScMn$_6$(Sn$_{0.78}$Ga$_{0.22}$)$_{6}$, collected at room temperature. To confirm the crystal structure, we performed Rietveld refinement on the XRD patterns using the Fullprof software \cite{Carvaja}. The room-temperature XRD patterns are shown in Fig. S1. The close match between the observed data (symbols) and the refined data (lines) indicates that the compound crystallizes in the hexagonal structure with space group P6/mmm (No. 191). A schematic depicting the crystallographic axes used for applying the current, external magnetic field, and measuring the voltage is presented in Fig. S3.\\

\noindent\textbf{3. Fermi surface of ScMn$_6$(Sn$_{0.78}$Ga$_{0.22}$)$_{6}$.}\\
Figure S4(a-c) presents the Fermi surface (FS) map obtained using ARPES with a photon energy of 110 eV and linear horizontal (LH) polarized light. A circular Fermi pocket is observed at the $\overline{\Gamma}$ high-symmetry point, while a hexagonal pocket is seen along the $\overline{\Gamma}$–$\overline{\text{M}}$/$\overline{\text{K}}$ high-symmetry line. Additionally, intense pockets are found at the $\overline{\text{K}}$ point, and the Brillouin zone (BZ) is outlined by blue dotted lines. As the binding energy increases, the size of the hexagonal and circular pockets at the $\overline{\Gamma}$ and $\overline{\Gamma}$–$\overline{\text{M}}$/$\overline{\text{K}}$ lines decreases, indicating their electron-like nature. Conversely, the pockets at the $\overline{\text{K}}$ point grow in size, showing hole-like behavior. These findings are consistent with the FS analysis discussed in the main manuscript. The FS calculated using DFT is presented in Fig. S5(a). The key features of the experimental FS, such as the circular pocket at $\overline{\Gamma}$, the hexagonal pocket along the $\overline{\Gamma}$–$\overline{\text{M}}$/$\overline{\text{K}}$ line, and the pockets at the $\overline{\text{K}}$ point, align well with the DFT results. As binding energy increases, see Fig. S5(b), the shrinking of the circular and hexagonal pockets at $\overline{\Gamma}$ and the expansion of the $\overline{\text{K}}$ pocket indicate electron-like and hole-like behavior, respectively, in excellent agreement with the ARPES-measured FS.\\ 

\begin{figure} 
	\includegraphics[width=8cm]{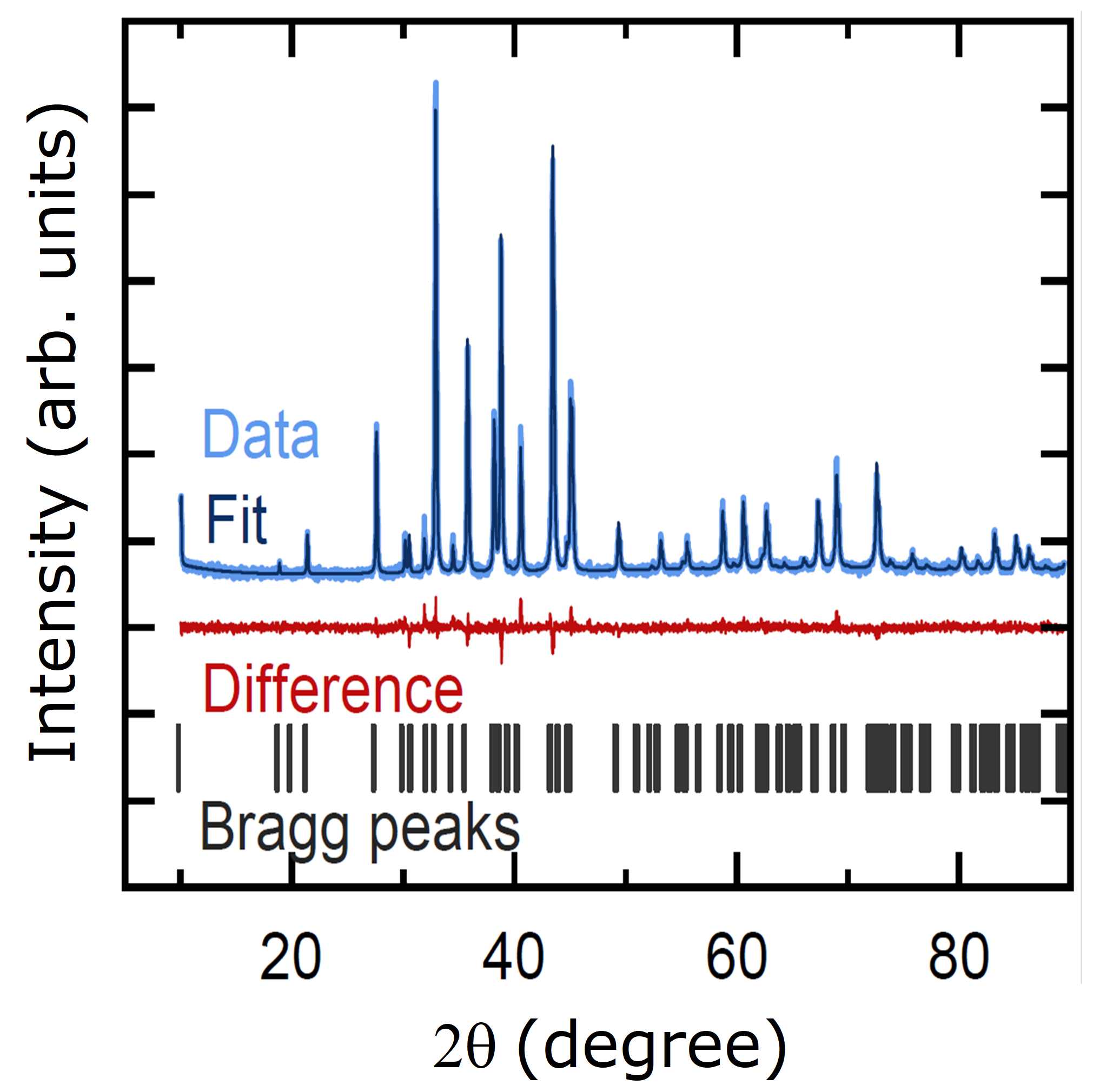}
     \vspace{-1ex}
	\caption{X-ray diffraction data. Rietveld refinement plot of ScMn$_6$(Sn$_{0.78}$Ga$_{0.22}$)$_{6}$ at room temperature. The experimental data are represented by blue colors and the solid black line represents the simulated XRD data. The trace in the middle (red line) is the plot of the difference between the observed and the calculated pattern. The black vertical lines at the bottom show the positions calculated for Bragg reflection.} 
\label{fig2}
\end{figure}

\begin{figure} 
	\includegraphics[width=10cm]{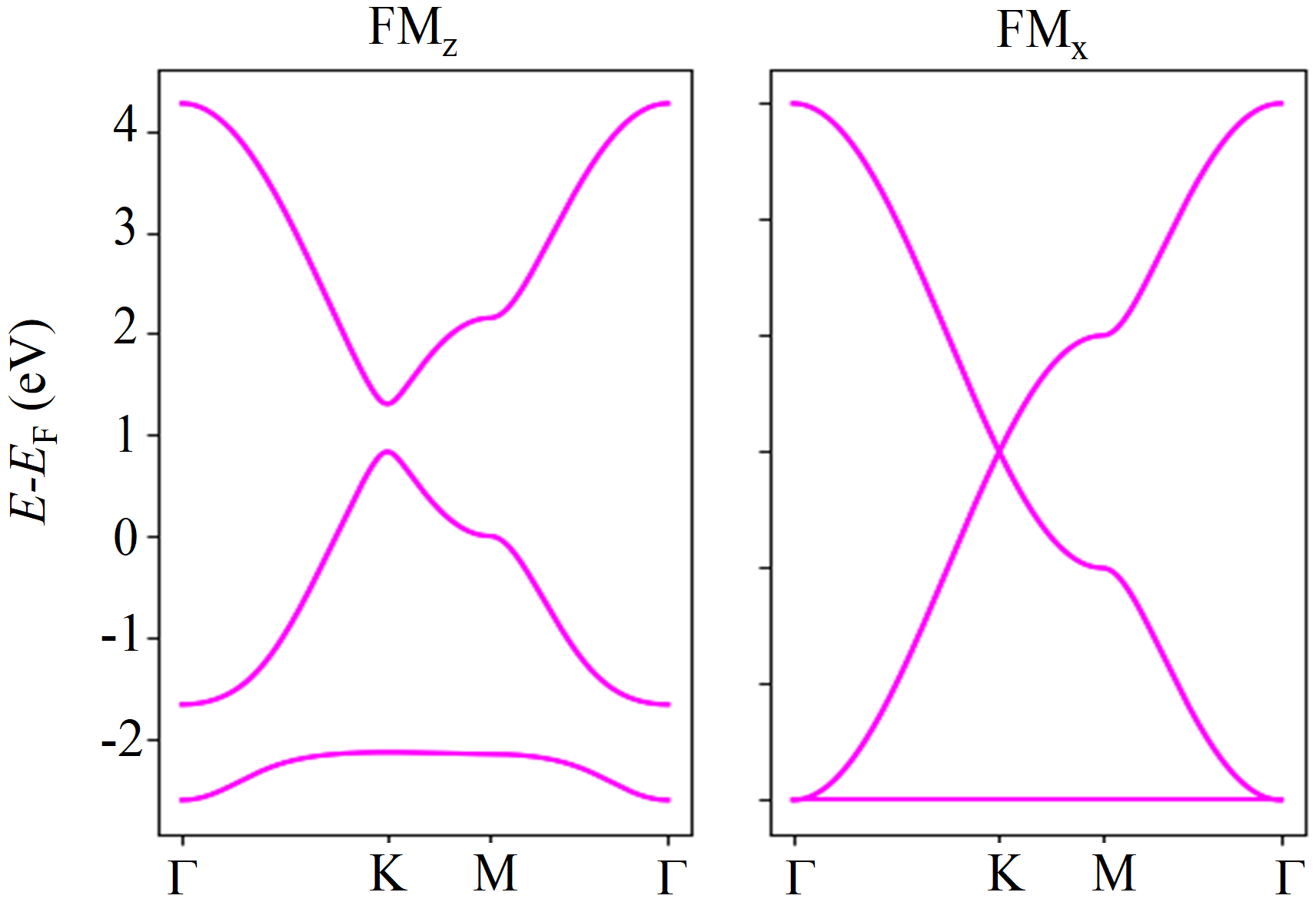} 
       \vspace{-1ex}
	\caption{Effect of the orientation of the magnetization axis on the Dirac cone. A simple tight-binding model illustrates the electronic band structures of the kagome lattice. The left panel shows the band structure when the magnetization is aligned along the \textit{c}-axis, while the right panel presents the band structure when the magnetization is oriented along the \textit{ab}-plane. This comparison highlights how the orientation of the magnetization axis impacts the Dirac cone and the electronic properties of the kagome lattice.}
\label{fig3}
\end{figure}

\begin{figure} 
	\includegraphics[width=8cm]{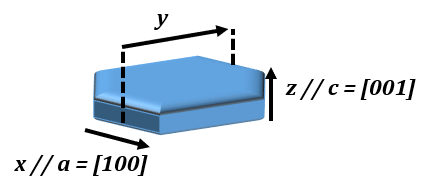}
     \vspace{-1ex}
	\caption{The schematic above shows the directions of the applied electrical current, external magnetic field and the measured voltage.} 
\label{fig2}
\end{figure}

\begin{figure} 
	\includegraphics[width=15cm]{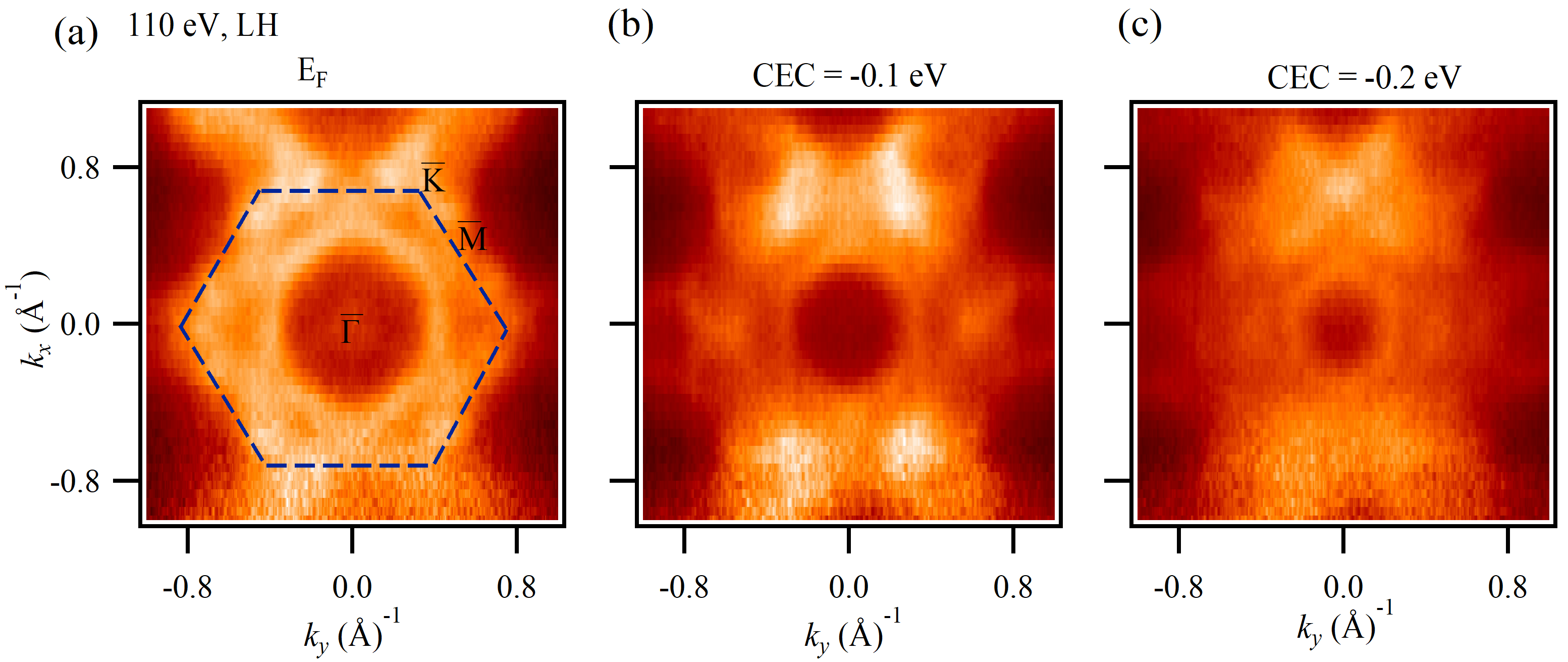}
    \vspace{-1ex}
	\caption{ARPES measured FS. (a) The Fermi surface (FS) measured using ARPES with a photon energy of 110 eV and linearly horizontally (LH) polarized light. Key high-symmetry points are indicated. (b-c) Constant energy contours measured at binding energies of -0.1 eV and -0.2 eV, respectively. These contours provide insights into the evolution of the electronic structure with increasing binding energy.}
\label{fig2}
\end{figure}

\begin{figure} 
	\includegraphics[width=12cm]{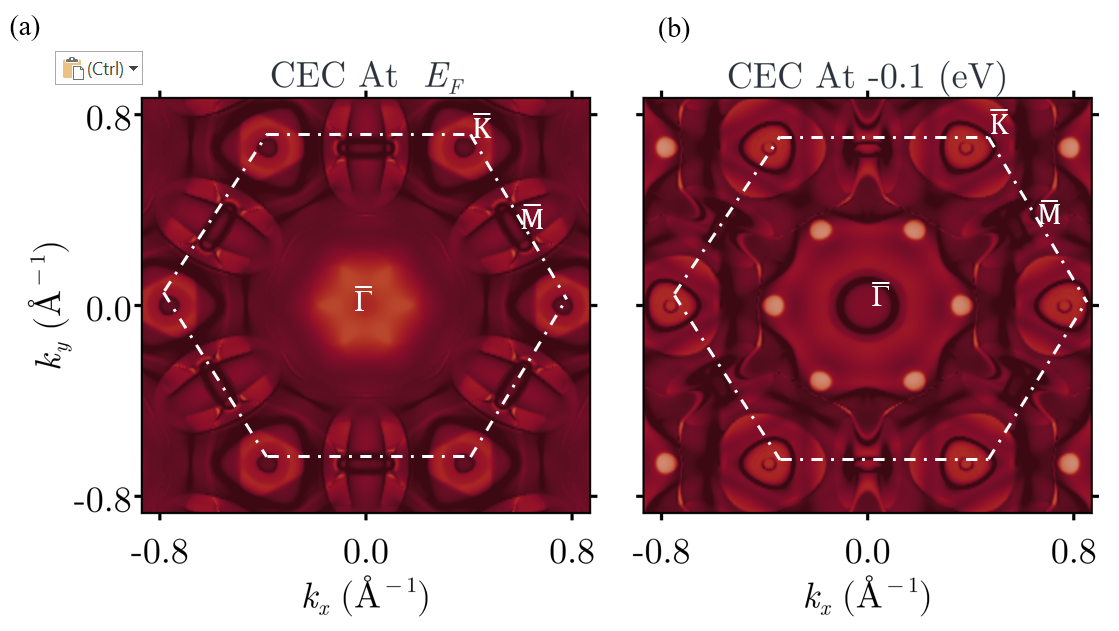}
    \vspace{-1ex}
	\caption{DFT calculated FS. (a) Projected bulk Fermi surface obtained through DFT calculations. Key high-symmetry points are marked for reference. (b) Constant energy contour obtained at a binding energy of -0.1 eV, illustrating the electronic structure at this specific energy level.}
\label{fig2}
\end{figure}

\noindent\textbf{4. Observation of Dirac cone from ARPES and DFT calculations.}\\
The DFT-calculated band dispersion along the $\text{K}$–$\Gamma$–$\text{K}$ high-symmetry line presented in Fig. S6 indicates the presence of a Dirac cone at the $\overline{\text{K}}$ point. This Dirac cone is highlighted with red circles and marked by black arrows. The observation of the Dirac cone at the $\text{K}$ point, positioned below the Fermi level (E$_F$), aligns well with the ARPES-measured band dispersion. The ARPES-measured band dispersion, taken using LH polarization along the $\overline{\Gamma}$–$\overline{\text{K}}$–$\overline{\text{M}}$–$\overline{\text{K}}$–$\overline{\Gamma}$ high-symmetry line, is shown in Fig. S7(a-e), where the Dirac cone at the $\overline{\text{K}}$ point below E$_F$ is clearly visible. Green arrows indicate the Dirac cone, with white dashed lines added for visual clarity. Additionally, the DFT-calculated band dispersion along the M–K–$\Gamma$ high-symmetry line as shown in Fig. S7(f) further corroborates the presence of the Dirac cone below E$_F$, reinforcing the consistency with our ARPES data.\\ 

\begin{figure} 
	\includegraphics[width=6cm]{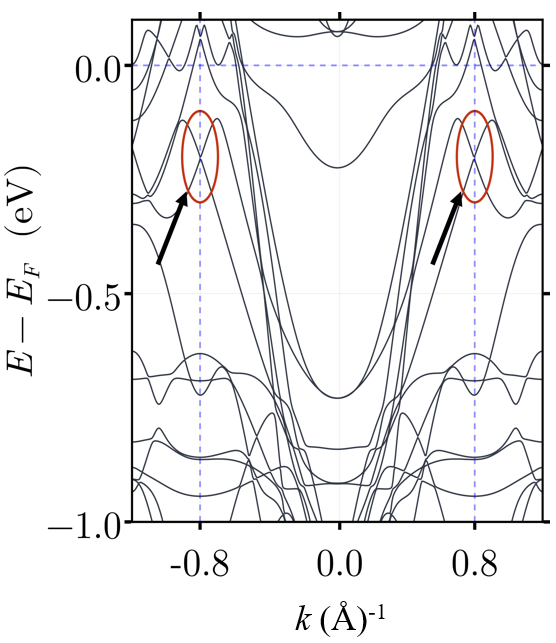}
    \vspace{-1ex}
	\caption{Observation of the Dirac cone from DFT calculations. DFT-calculated band dispersion along the $\overline{{\text{K}}}$--$\overline{\Gamma}$--$\overline{{\text{K}}}$ high-symmetry line.}
\label{fig2}
\end{figure}

\begin{figure} 
	\includegraphics[width=15cm]{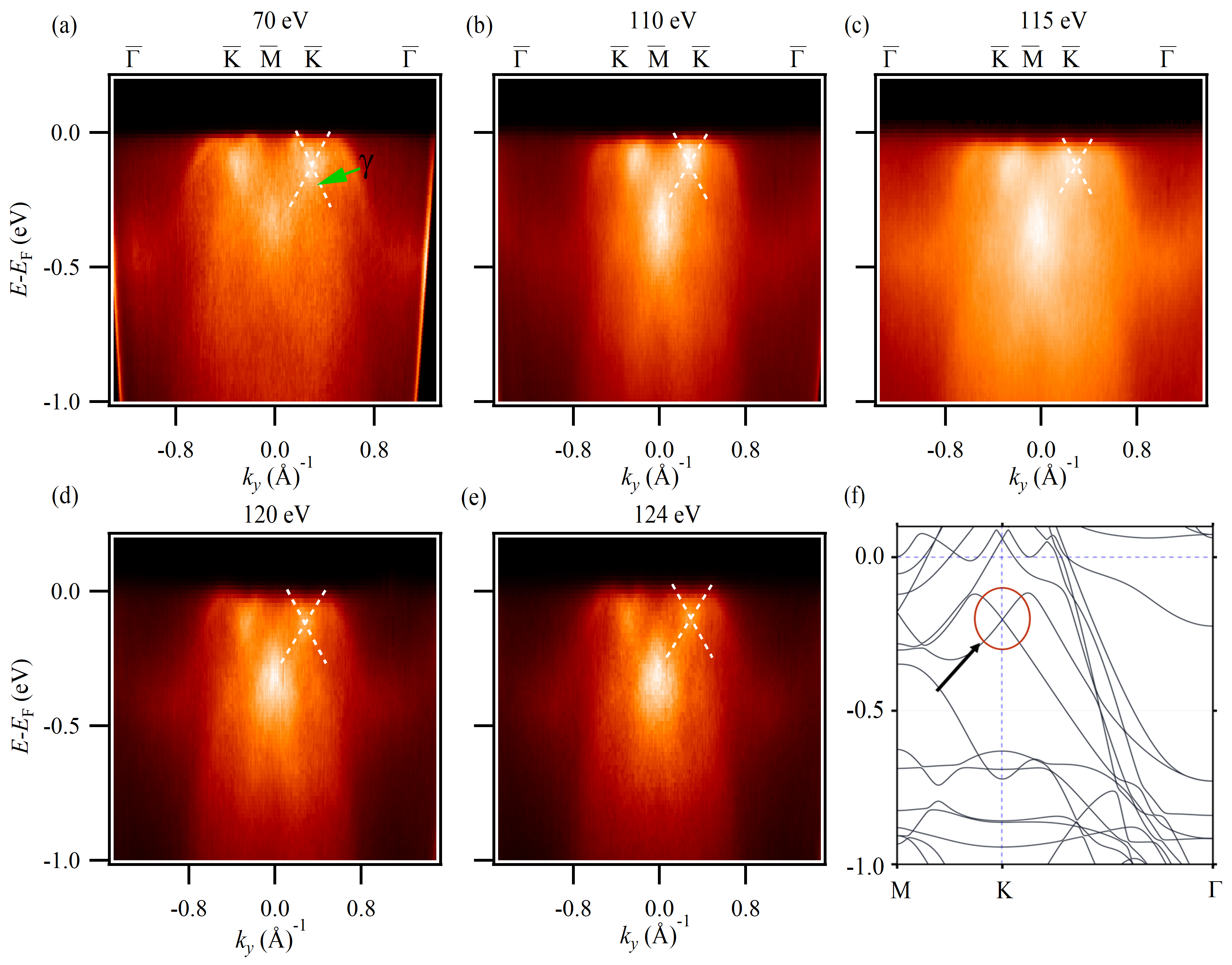}
    \vspace{-1ex}
	\caption{Observation of the Dirac cone from photon-energy dependent ARPES measurements and DFT calculations. (a-e) ARPES measured band-dispersion presented along the $\overline{\Gamma}$--$\overline{{\text{K}}}$--$\overline{{\text{M}}}$--$\overline{\text{K}}$--$\overline{\Gamma}$ high-symmetry line at various photon energies, as indicated at the top of the plots using LH polarization. The Dirac cone is indicated by green arrow and white-dashed lines. (f) DFT calculated band-dispersion along the M--K--$\Gamma$ high-symmetry line, with black arrow indicating the presence of the Dirac cone.}
\label{fig2}
\end{figure}

\noindent\textbf{5. Observation of flat band from ARPES measurements.}\\
As discussed in the main text, a flat band is clearly observed in Fig. S8(a-d) at a binding energy of approximately $\sim$ -0.4 eV, marked by a blue arrow and labeled as $\alpha$. This flat band persists across all photon energy measurements, although its intensity varies. The corresponding density of states can be visualized in the momentum-integrated EDCs of Fig. S8(a-d), which exhibit distinct peaks arising from the flat-band. The variation in intensity may be attributed to changes in the photoemission cross section, a well-known effect in ARPES measurements. To quantitatively assess the dispersion of this flat band, energy distribution curves (EDCs) were analyzed at successive momenta, and the peak positions were extracted. The fitted peak energies are overlaid on the ARPES intensity map in Fig. S9(b) as green symbols, with the associated uncertainties reflecting the fitting procedure and experimental energy resolution. The extracted dispersion shows only weak energy variation across momentum and remains within the uncertainty limits, indicating a minimal bandwidth. This analysis confirms the nearly dispersionless character of the band.\\

\begin{figure} 
	\includegraphics[width=15cm]{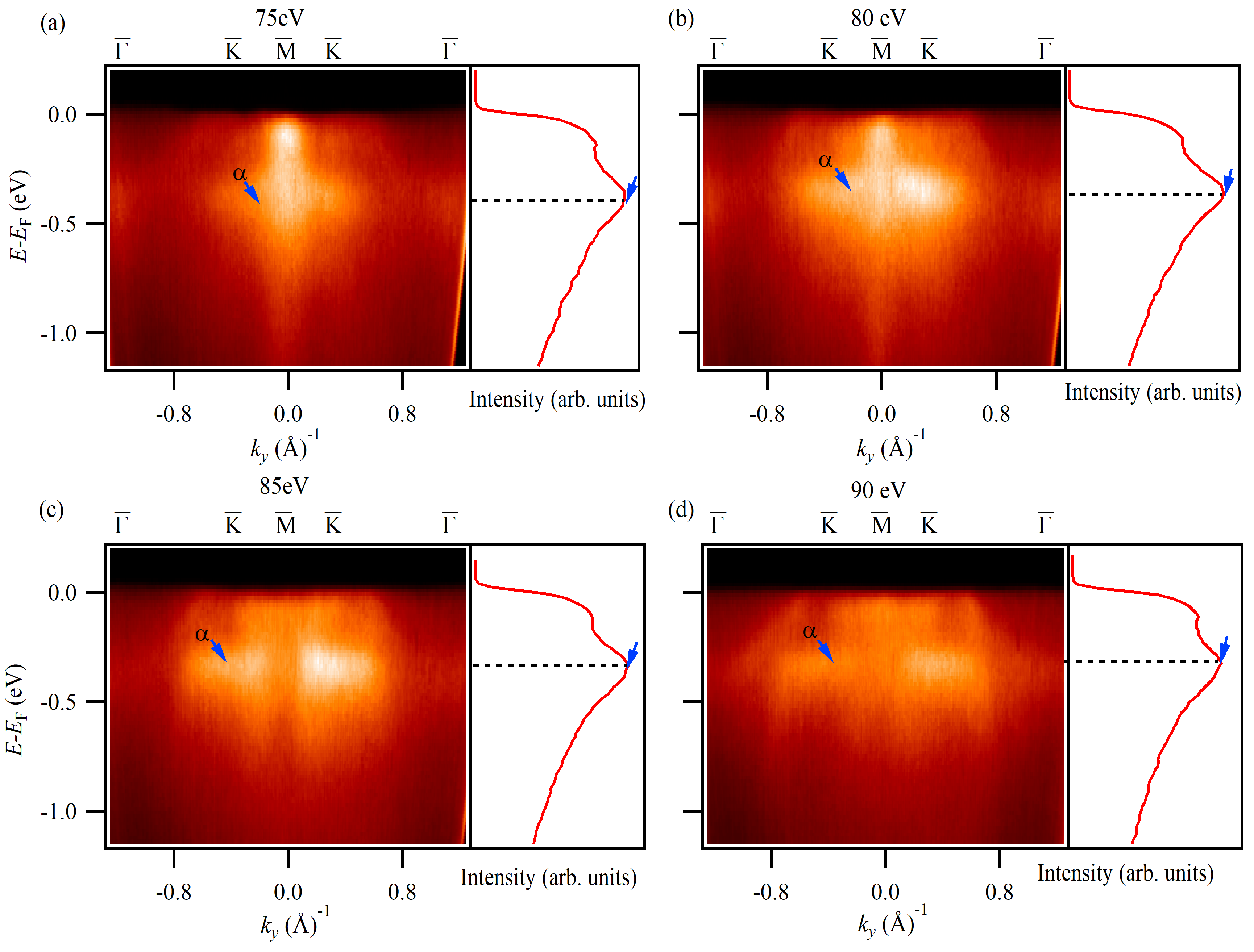}
     \vspace{-1ex}
	\caption{Observation of flat band. (a-d) ARPES-measured band dispersion and the corresponding integrated energy distribution curves (EDCs) within a momentum window of (-0.8 \AA$^{-1}$ to 0 \AA$^{-1}$) along the $\overline{\Gamma}$--$\overline{\text{K}}$--$\overline{\text{M}}$--$\overline{\text{K}}$--$\overline{\Gamma}$ high-symmetry line, measured using various photon energies as indicated at the top of the plots with LV polarization. The flat band is marked by a blue arrow and denoted by the symbol $\alpha$.}
\label{fig2}
\end{figure}

\begin{figure} 
	\includegraphics[width=15cm]{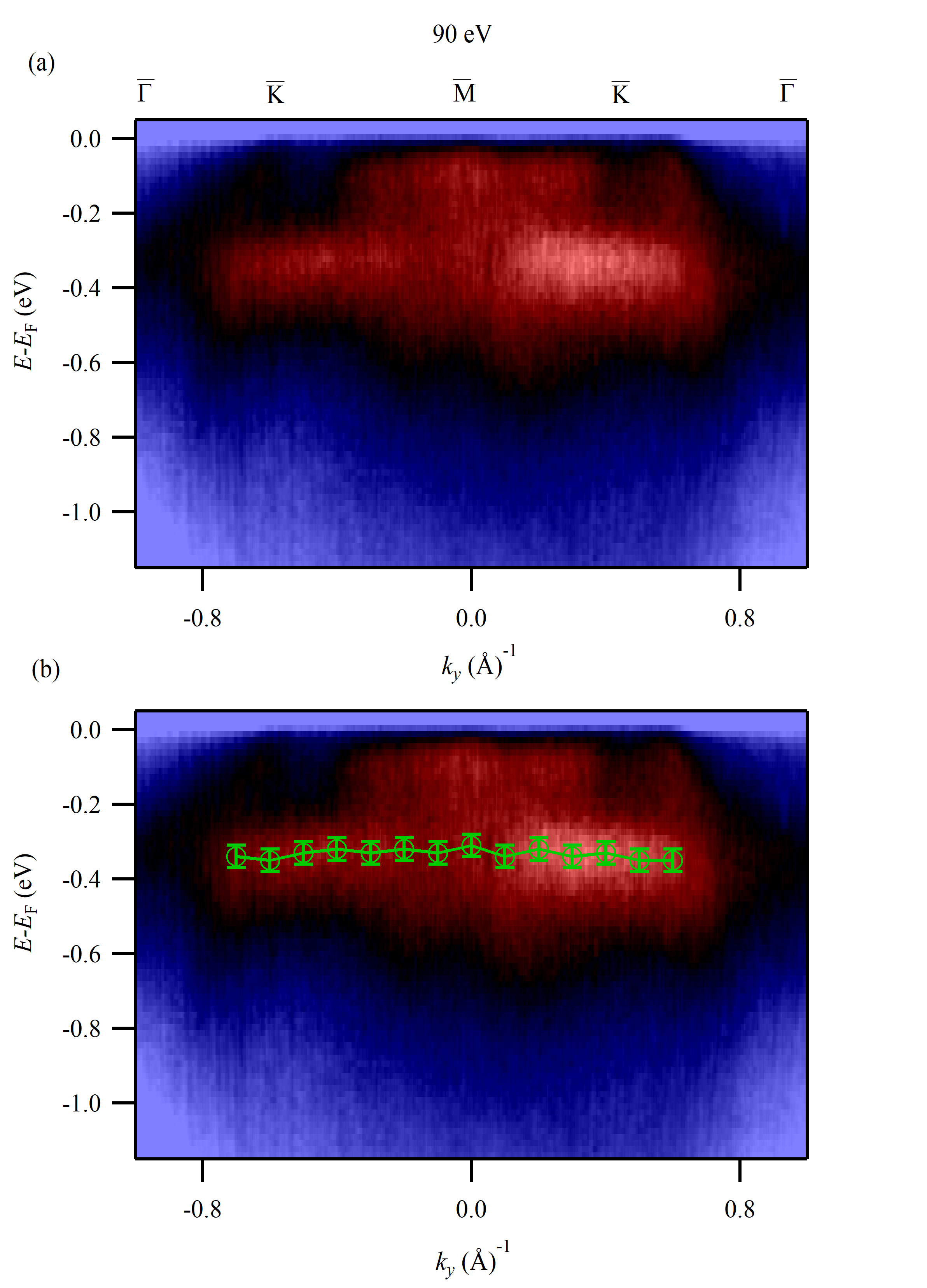}
     \vspace{-1ex}
	\caption{(a) ARPES intensity map along the $\overline{\Gamma}$--$\overline{\text{K}}$--$\overline{\text{M}}$--$\overline{\text{K}}$--$\overline{\Gamma}$ direction. (b) ARPES intensity map along the $\overline{\Gamma}$--$\overline{\text{K}}$--$\overline{\text{M}}$--$\overline{\text{K}}$--$\overline{\Gamma}$ direction with the overlaid curve showing the flat band dispersion obtained by extracting EDCs from the ARPES map within a momentum range of -0.6 \AA$^{-1}$ to 0.6 \AA$^{-1}$, where the flat band feature is clearly visible. Peak positions were determined by fitting the EDCs with Gaussian functions. Error bars of $\pm$ 25 meV represent the uncertainty in determining the band energy. The flatness of this band indicates very low dispersion and a high effective mass of the carriers.}
\label{fig2}
\end{figure}

\clearpage
\end{document}